\RequirePackage{lineno}
\documentclass[aps,prl,twocolumn,superscriptaddress,showpacs,preprintnumbers,amsmath,amssymb,doublespace]{revtex4}
\usepackage{fancyhdr}
\usepackage{lineno}
\usepackage{setspace}
\usepackage{slashed}

\usepackage{graphicx}% Include figure files
\usepackage{dcolumn}% Align table columns on decimal point
\usepackage{bm}% bold math
\usepackage{longtable}

\begin{document}
%\preprint{CDF/DOC/TOP/CDFR/9036}
%\pagewiselinenumbers
%\doublespace

%%%%%%%%%%%%%%%%%%%%%%%%%%%%%%THE NICE REVTEK WAY%%%%%%%%%%%%%%%%%%%%%%%%%%%%%%%%
\title{\boldmath Observation of the Production of a $W$ Boson in Association with a Single Charm Quark} 

%\author{Lucio Cerrito, John Paul Chou, Melissa Franklin, Shulamit Moed Sher}
%\author{the CDF Collaboration}
%\affiliation{(for the CDF Collaboration)}

\date{\today}
\begin{abstract}  
The first observation of the production of a $W$ boson with a single charm quark ($c$) jet in $p\bar{p}$ collisions at $\sqrt{s}=1.96~{\rm TeV}$ is reported. The analysis uses data corresponding to $4.3$~fb$^{-1}$, recorded with the CDF~II detector at the Fermilab Tevatron. Charm quark candidates are selected through the identification of an electron or muon from charm-hadron semileptonic decay within a hadronic jet, and a $Wc$ signal is observed with a significance of 5.7 standard deviations. The production cross section $\sigma_{Wc}\left(p_{Tc}>20~{\rm GeV}/c,~|\eta_c|<1.5\right)\times B(W\rightarrow \ell\nu)$ is measured to be $13.6^{+3.4}_{-3.1}~{\rm pb}$ and is in agreement with theoretical expectations.  
From this result the magnitude of the quark-mixing matrix element $V_{cs}$ is derived, $|V_{cs}|=1.08\pm0.16$ along with a lower limit of $|V_{cs}|>0.71$ at the 95\% confidence level, assuming that the $Wc$ production through $c$ to $s$ quark coupling is dominant.

\end{abstract}
\affiliation{Institute of Physics, Academia Sinica, Taipei, Taiwan 11529, Republic of China}
\affiliation{Argonne National Laboratory, Argonne, Illinois 60439, USA}
\affiliation{University of Athens, 157 71 Athens, Greece}
\affiliation{Institut de Fisica d'Altes Energies, ICREA, Universitat Autonoma de Barcelona, E-08193, Bellaterra (Barcelona), Spain}
\affiliation{Baylor University, Waco, Texas 76798, USA}
\affiliation{Istituto Nazionale di Fisica Nucleare Bologna, $^{ee}$University of Bologna, I-40127 Bologna, Italy}
\affiliation{University of California, Davis, Davis, California 95616, USA}
\affiliation{University of California, Los Angeles, Los Angeles, California 90024, USA}
\affiliation{Instituto de Fisica de Cantabria, CSIC-University of Cantabria, 39005 Santander, Spain}
\affiliation{Carnegie Mellon University, Pittsburgh, Pennsylvania 15213, USA}
\affiliation{Enrico Fermi Institute, University of Chicago, Chicago, Illinois 60637, USA}
\affiliation{Comenius University, 842 48 Bratislava, Slovakia; Institute of Experimental Physics, 040 01 Kosice, Slovakia}
\affiliation{Joint Institute for Nuclear Research, RU-141980 Dubna, Russia}
\affiliation{Duke University, Durham, North Carolina 27708, USA}
\affiliation{Fermi National Accelerator Laboratory, Batavia, Illinois 60510, USA}
\affiliation{University of Florida, Gainesville, Florida 32611, USA}
\affiliation{Laboratori Nazionali di Frascati, Istituto Nazionale di Fisica Nucleare, I-00044 Frascati, Italy}
\affiliation{University of Geneva, CH-1211 Geneva 4, Switzerland}
\affiliation{Glasgow University, Glasgow G12 8QQ, United Kingdom}
\affiliation{Harvard University, Cambridge, Massachusetts 02138, USA}
\affiliation{Division of High Energy Physics, Department of Physics, University of Helsinki and Helsinki Institute of Physics, FIN-00014, Helsinki, Finland}
\affiliation{University of Illinois, Urbana, Illinois 61801, USA}
\affiliation{The Johns Hopkins University, Baltimore, Maryland 21218, USA}
\affiliation{Institut f\"{u}r Experimentelle Kernphysik, Karlsruhe Institute of Technology, D-76131 Karlsruhe, Germany}
\affiliation{Center for High Energy Physics: Kyungpook National University, Daegu 702-701, Korea; Seoul National University, Seoul 151-742, Korea; Sungkyunkwan University, Suwon 440-746, Korea; Korea Institute of Science and Technology Information, Daejeon 305-806, Korea; Chonnam National University, Gwangju 500-757, Korea; Chonbuk National University, Jeonju 561-756, Korea}
\affiliation{Ernest Orlando Lawrence Berkeley National Laboratory, Berkeley, California 94720, USA}
\affiliation{University of Liverpool, Liverpool L69 7ZE, United Kingdom}
\affiliation{University College London, London WC1E 6BT, United Kingdom}
\affiliation{Centro de Investigaciones Energeticas Medioambientales y Tecnologicas, E-28040 Madrid, Spain}
\affiliation{Massachusetts Institute of Technology, Cambridge, Massachusetts 02139, USA}
\affiliation{Institute of Particle Physics: McGill University, Montr\'{e}al, Qu\'{e}bec H3A~2T8, Canada; Simon Fraser University, Burnaby, British Columbia V5A~1S6, Canada; University of Toronto, Toronto, Ontario M5S~1A7, Canada; and TRIUMF, Vancouver, British Columbia V6T~2A3, Canada}
\affiliation{University of Michigan, Ann Arbor, Michigan 48109, USA}
\affiliation{Michigan State University, East Lansing, Michigan 48824, USA}
\affiliation{Institution for Theoretical and Experimental Physics, ITEP, Moscow 117259, Russia}
\affiliation{University of New Mexico, Albuquerque, New Mexico 87131, USA}
\affiliation{The Ohio State University, Columbus, Ohio 43210, USA}
\affiliation{Okayama University, Okayama 700-8530, Japan}
\affiliation{Osaka City University, Osaka 588, Japan}
\affiliation{University of Oxford, Oxford OX1 3RH, United Kingdom}
\affiliation{Istituto Nazionale di Fisica Nucleare, Sezione di Padova-Trento, $^{ff}$University of Padova, I-35131 Padova, Italy}
\affiliation{University of Pennsylvania, Philadelphia, Pennsylvania 19104, USA}
\affiliation{Istituto Nazionale di Fisica Nucleare Pisa, $^{gg}$University of Pisa, $^{hh}$University of Siena and $^{ii}$Scuola Normale Superiore, I-56127 Pisa, Italy, $^{mm}$INFN Pavia and University of Pavia, I-27100 Pavia, Italy}
\affiliation{University of Pittsburgh, Pittsburgh, Pennsylvania 15260, USA}
\affiliation{Purdue University, West Lafayette, Indiana 47907, USA}
\affiliation{University of Rochester, Rochester, New York 14627, USA}
\affiliation{The Rockefeller University, New York, New York 10065, USA}
\affiliation{Istituto Nazionale di Fisica Nucleare, Sezione di Roma 1, $^{jj}$Sapienza Universit\`{a} di Roma, I-00185 Roma, Italy}
\affiliation{Texas A\&M University, College Station, Texas 77843, USA}
\affiliation{Istituto Nazionale di Fisica Nucleare Trieste/Udine, I-34100 Trieste, Italy; $^{kk}$University of Udine, I-33100 Udine, Italy}
\affiliation{University of Tsukuba, Tsukuba, Ibaraki 305, Japan}
\affiliation{Tufts University, Medford, Massachusetts 02155, USA}
\affiliation{University of Virginia, Charlottesville, Virginia 22906, USA}
\affiliation{Waseda University, Tokyo 169, Japan}
\affiliation{Wayne State University, Detroit, Michigan 48201, USA}
\affiliation{University of Wisconsin, Madison, Wisconsin 53706, USA}
\affiliation{Yale University, New Haven, Connecticut 06520, USA}

\author{T.~Aaltonen}
\affiliation{Division of High Energy Physics, Department of Physics, University of Helsinki and Helsinki Institute of Physics, FIN-00014, Helsinki, Finland}
\author{S.~Amerio}
\affiliation{Istituto Nazionale di Fisica Nucleare, Sezione di Padova-Trento, $^{ff}$University of Padova, I-35131 Padova, Italy}
\author{D.~Amidei}
\affiliation{University of Michigan, Ann Arbor, Michigan 48109, USA}
\author{A.~Anastassov$^x$}
\affiliation{Fermi National Accelerator Laboratory, Batavia, Illinois 60510, USA}
\author{A.~Annovi}
\affiliation{Laboratori Nazionali di Frascati, Istituto Nazionale di Fisica Nucleare, I-00044 Frascati, Italy}
\author{J.~Antos}
\affiliation{Comenius University, 842 48 Bratislava, Slovakia; Institute of Experimental Physics, 040 01 Kosice, Slovakia}
\author{G.~Apollinari}
\affiliation{Fermi National Accelerator Laboratory, Batavia, Illinois 60510, USA}
\author{J.A.~Appel}
\affiliation{Fermi National Accelerator Laboratory, Batavia, Illinois 60510, USA}
\author{T.~Arisawa}
\affiliation{Waseda University, Tokyo 169, Japan}
\author{A.~Artikov}
\affiliation{Joint Institute for Nuclear Research, RU-141980 Dubna, Russia}
% ...

\collaboration{CDF Collaboration\footnote{With visitors from
$^a$Istituto Nazionale di Fisica Nucleare, Sezione di Cagliari, 09042 Monserrato (Cagliari), Italy,
$^b$University of California Irvine, Irvine, CA 92697, USA,
$^c$University of California Santa Barbara, Santa Barbara, CA 93106, USA,
$^d$University of California Santa Cruz, Santa Cruz, CA 95064, USA,
$^e$Institute of Physics, Academy of Sciences of the Czech Republic, 182~21, Czech Republic,
$^f$CERN, CH-1211 Geneva, Switzerland,
$^g$Cornell University, Ithaca, NY 14853, USA,
$^h$University of Cyprus, Nicosia CY-1678, Cyprus,
$^i$Office of Science, U.S. Department of Energy, Washington, DC 20585, USA,
$^j$University College Dublin, Dublin 4, Ireland,
$^k$ETH, 8092 Zurich, Switzerland,
$^l$University of Fukui, Fukui City, Fukui Prefecture, Japan 910-0017,
$^m$Universidad Iberoamericana, Lomas de Santa Fe, M\'{e}xico, C.P. 01219, Distrito Federal,
$^n$University of Iowa, Iowa City, IA 52242, USA,
$^o$Kinki University, Higashi-Osaka City, Japan 577-8502,
$^p$Kansas State University, Manhattan, KS 66506, USA,
$^q$Ewha Womans University, Seoul, 120-750, Korea,
$^r$University of Manchester, Manchester M13 9PL, United Kingdom,
$^s$Queen Mary, University of London, London, E1 4NS, United Kingdom,
$^t$University of Melbourne, Victoria 3010, Australia,
$^u$Muons, Inc., Batavia, IL 60510, USA,
$^v$Nagasaki Institute of Applied Science, Nagasaki 851-0193, Japan,
$^w$National Research Nuclear University, Moscow 115409, Russia,
$^x$Northwestern University, Evanston, IL 60208, USA,
$^y$University of Notre Dame, Notre Dame, IN 46556, USA,
$^z$Universidad de Oviedo, E-33007 Oviedo, Spain,
$^{aa}$CNRS-IN2P3, Paris, F-75205 France,
$^{bb}$Texas Tech University, Lubbock, TX 79609, USA,
$^{cc}$Universidad Tecnica Federico Santa Maria, 110v Valparaiso, Chile,
$^{dd}$Yarmouk University, Irbid 211-63, Jordan,
$^{ll}$Universite catholique de Louvain, 1348 Louvain-La-Neuve, Belgium 
}}
\noaffiliation

\maketitle
%%%%%%%%%%%%%%%%%%%%%%%%%%%%%%%%%%%%%%%%%%%%%%%%%%%%%%%%%%%%%%%%%%%%%%%%%%%%%%%%%%%%%%
               
The associated production of the $W$ boson with a single charm quark in proton-antiproton collisions is described at lowest order in the standard model (SM) by quark-gluon fusion ($gq\rightarrow Wc$), where $q$ denotes a $d$, $s$, or $b$ quark. At the Tevatron proton-antiproton collider, the larger $d$ quark parton distribution function (PDF) in the proton is compensated by the small quark-mixing (Cabibbo-Kobayashi-Maskawa or CKM) matrix element $|V_{cd}|$, so that only about 20\% of the total $Wc$ production rate is due to $gd\rightarrow Wc$, with the majority due to strange quark-gluon fusion. The contribution from $gb\rightarrow Wc$ is also heavily suppressed by $|V_{cb}|$ and the $b$ quark PDF. The $Wc$ production cross section is therefore particularly sensitive to the gluon and $s$ quark PDFs \cite{mangano,lai}, at a momentum transfer $Q^2$ of the order of the $W$ boson mass ($M_W$), and to the magnitude of the CKM matrix element $V_{cs}$. Measurements of $Wc$ production in high energy $p\bar{p}$ collisions are of interest because they constrain the proton's $s$ quark PDF at momentum transfers about three orders of magnitude higher than in neutrino-nucleon scattering \cite{nutev}. Finally, the $Wc$ final state is similar to final state of other processes, such as single top-quark production, neutral and charged Higgs boson production, and supersymmetric top-quark production. The techniques developed here could lead to a better understanding of those samples and their searches. Calculations of $W$ + heavy~quark production are available at leading order (LO) and next-to-leading order (NLO) in quantum chromodynamics (QCD) \cite{keller}, with the NLO cross section prediction about 50\% larger than the LO calculation. Overall, the uncertainty on the NLO theoretical expectation for the $Wc$ production cross section at the Tevatron is 10\%--20\%, depending on the charm phase space considered.

We present the first observation of $p\bar{p}\rightarrow Wc$ production. The charm quark is identified through the semileptonic decay of the charm hadron into an electron or muon (referred to in this Letter as \textquotedblleft soft leptons\textquotedblright). This measurement supersedes our previous result \cite{cdfwc1}, where the cross section for $p\bar{p}\rightarrow Wc$ was determined with a precision of approximately 30\% and a statistical significance of about 3 standard deviations. The present analysis is performed using a data set more than twice as large and signal events with soft electrons are included to increase the acceptance, leading to a final signal sample about 3 times larger than in the previous publication. The analysis exploits the correlation between the charge of the $W$ boson and the charge of the soft lepton from the semileptonic decay of the charm hadron. Charge conservation in the process  $gq\rightarrow Wc$ ($q=d,s$) allows only $W^+\bar{c}$ and $W^-c$ final states; as a result the charge of the lepton from the semileptonic decay of the $c$ quark and the charge of the $W$ boson are always of opposite sign, neglecting any effects due to slow-rate charm quark oscillations \cite{pdg}. 

The $W$ boson is identified through its leptonic decay by looking for an isolated electron (muon) carrying large transverse energy $E_T$ (momentum $p_T$), with respect to the beam line. The neutrino escapes the detector, causing an imbalance of total transverse energy, referred to as \textquotedblleft missing $E_T$\textquotedblright ($\slashed{E}_T$) \cite{cdfsys}. Quarks hadronize and are observed as jets of charged and neutral particles. Charm jets are identified by requiring an electron or muon candidate within the jet (\textquotedblleft soft lepton tagging\textquotedblright or  \textquotedblleft SLT$_\ell$\textquotedblright). Events are classified based on whether the charge of the lepton from the $W$ boson and the charge of the soft lepton are of opposite sign (OS) or same sign (SS). The $Wc$ production cross section is then calculated using the formula
\begin{equation}\label{eq:Xsec}
\sigma_{Wc}= \frac{N_{\rm tot}^{\rm OS-SS}-N^{\rm OS-SS}_{\rm bkg}}{S~ A \int L\;dt},
\end{equation}
where $N^{\rm OS-SS}_{\rm tot}$ ($N^{\rm OS-SS}_{\rm bkg}$) is the difference in the number of OS and SS events in data (background), $A$ is the product of the efficiency, for identifying $Wc$ events, with the kinematical and geometrical acceptance, and $\int L\;dt$ is the integrated luminosity of the data sample. The quantity $S=(N_{Wc}^{\rm OS}-N_{Wc}^{\rm SS})/(N_{Wc}^{\rm OS}+N_{Wc}^{\rm SS})$ accounts for the charge asymmetry of the sample of real reconstructed $Wc$ events, which is less than unity due to dilution arising from hadronic decays in flight and hadrons misidentified as soft leptons. The terms $A$ and $S$, which are derived from a Monte Carlo (MC) simulation of $Wc$ events and the detector response, specify the unfolding from the observed same-sign subtracted $Wc$ event yield to the measured cross section. The cross section is defined through $A$ to correspond to the production of a $W$ boson over the entire kinematic range associated with a single charm quark with $p_{Tc}>20~{\rm GeV}/c$, $|\eta_c|<1.5$. The phase space of the charm is restricted to approximately match the detector acceptance of the charm quark, which minimizes the theoretical uncertainties on $A$.  In the determination of $A$, the $Wc$ signal is defined to include events with a single charm quark and allows for additional jets; contributions from all sources of $W$ bosons associated with $c\bar{c}$ pairs are not considered in the acceptance since they cancel out in the same-sign subtraction, owing to the largely charge-symmetric detector response.

The CDF~II detector is described in detail elsewhere \cite{CDF}. The data sample, produced in $p\bar{p}$ collisions at $\sqrt{s}=1.96~{\rm TeV}$ during Run$~$II of the Fermilab Tevatron, corresponds to $4.3\pm0.3~{\rm fb}^{-1}$ and was collected between March 2002 and March 2009. Events are selected with an inclusive-lepton online event selection (trigger) requiring an electron (muon) with $E_T>18~{\rm GeV}$ ($p_T>18~{\rm GeV}/c$) \cite{smtprd2}. Further selection requires exactly one isolated electron (muon), both with isolation parameter $I<0.1$ \cite{iso}, with $E_T$ ($p_T$) greater than 20~GeV (20~GeV/$c$) and $|\eta|<1.1$. The event must also have $\slashed{E}_T>25$ GeV and exactly one jet with $E_T>20~{\rm GeV}$ and $|\eta|<2.0$. The transverse mass of the $W$ boson candidates is required to be greater than 20~GeV$/c^2$ \cite{mtw}. Jets are identified using a fixed-cone algorithm with a cone opening of $\Delta R\equiv\sqrt{(\Delta\eta)^2+(\Delta\phi)^2}=0.4$ and are constrained to originate from the $p\bar{p}$ collision vertex. The jet energies are corrected for detector response, multiple interactions, and uninstrumented regions of the detector  \cite{jes}.

Muon candidates inside jets are identified by matching the trajectories of charged particles (tracks) of the jet, as measured in the inner tracking system, with track segments in the muon detectors. An SLT$_\mu$ \cite{smtprd1, smtprd2} must have $p_T>3$~GeV/$c$ and be within $\Delta R<0.6$ of a jet axis. Soft electrons from semileptonic heavy-flavor decay (SLT$_e$) are identified by tracks with $p_T>2$~GeV/$c$ that are associated with an electromagnetic shower in the central electromagnetic calorimeter, and must lie within $\Delta R<0.4$ of a jet axis. Furthermore, finely segmented wire and strip chambers are used to identify the collimated shower of the electron within the broader hadronic shower of the jet. Additional variables to discriminate soft electrons are based on the energy deposition, transverse shower shape, and track-shower distance~\cite{setprd, JPThesis}. To reduce background from dielectron and dimuon resonances, events are discarded where the invariant mass, computed from the same-flavor oppositely charged soft lepton and primary lepton, is consistent with $\Upsilon$ or $Z$ (for SLT$_\mu$), or greater than 45~GeV/$c^2$ (for SLT$_e$). Events are also discarded if the jet tagged by a soft muon has an electromagnetic fraction greater than 90\%, reducing the contamination from $Z\rightarrow \mu\mu$ decays with final-state radiation off one muon. To suppress QCD multijet background, we reject events for which the azimuthal angular difference between the $\slashed{E}_T$ and the jet is less than 0.3 rad. 

The dominant backgrounds to $Wc$ are due to the associated production of jets with the $W$ boson ($W$ + jets, excluding the $Wc$ under investigation), and from Drell-Yan production of $Z/\gamma^*$, with and without additional jets. Multijet QCD events and small contributions from diboson, single top, and $t\bar{t}$ production are also present. Backgrounds are estimated using a combination of MC simulation and control regions from the data. The MC simulations of $W$ + jets and $Z/\gamma^*$ + jets processes are performed using {\sc alpgen} ({\sc v2.1} \cite{alpgen}) interfaced with {\sc pythia} ({\sc v6.3} \cite{pythia}) for the parton shower (PS) evolution. The simulation of the $Wc$ signal is performed similarly and is referred to as LO + PS. Modeling of heavy-quark hadron decay is provided by {\sc evtgen} \cite{evtgen}. All samples are simulated using the {\sc CTEQ5L} PDF sets, with Tune {\sc BW} \cite{cteq5} to model the underlying event and the hadronization parameters. Events with a $Z\rightarrow\tau\tau$ decay are also simulated, as well as $Zb\bar{b}$ and $Zc\bar{c}$ final states. The production of $Z/\gamma^*$ + jets in the simulation is normalized by the measured exclusive $Z$ + 1 jet cross section \cite{zjets}.

The $W$ boson events that can mimic the $Wc$ signature consist of a $W$ boson associated with heavy-flavor quark pairs ($b\bar{b}$ and $c\bar{c}$) or light-flavor (LF) jets. However, since this measurement is sensitive to the excess of OS over SS events, such excess from $Wb\bar{b}$ and $Wc\bar{c}$ processes is negligible given the soft lepton can come from either the $b$ ($c$) or $\bar{b}$ ($\bar{c}$). On the other hand, $W$ + LF events enter the data sample when the jet is identified as a charm jet via a misreconstructed soft electron or soft muon tag (\textquotedblleft mistagging\textquotedblright). A small anticorrelation between the charge of the $W$ boson and the charge sign of the tracks in the jets recoiling against the $W$ is present, leading to a residual background contribution. We rely on a combination of MC simulations and data-driven techniques to estimate this contribution to the tagged sample: first the number of $W$ + jets events ($\simeq$97\% of which is $W$ + LF) is estimated in the sample of events before tagging the jet (\textquotedblleft pretag sample\textquotedblright) by subtracting from the data the initial pretag estimate of the signal and all other backgrounds. The number of tagged $W$ + jets events is obtained from this pretag estimate using a mistag probability parametrization \cite{smtprd2,setprd}. The probability of misidentifying a hadron as an SLT$_\ell$, denoted as the SLT$_\ell$ mistag probability, is parametrized as a function of the track curvature and $\eta$. The number of OS and SS events due to $W$ + LF are determined directly from the data by applying the mistag parametrizations to tracks in the $W$ + jet pretagged sample, and appropriately taking into account for the SLT$_e$ the contribution from photon conversions.

The second largest background to $Wc$ is due to the misreconstruction of $Z/\gamma^*$ + jets events. The two leptons from the $Z/\gamma^*$ decay can be misidentified as one lepton from a $W$ boson decay and one soft lepton, resulting in approximately 90\% charge asymmetry. These events are suppressed by the veto on the $Z$-mass region. Alternatively, only one lepton from the $Z$ boson decay is reconstructed in the event, which is typically assigned to be a $W$-decay lepton. In this case, the soft lepton results from the decay of heavy flavor or from the misreconstruction of a track from hadrons, and these events carry approximately 40\% asymmetry. The overall average charge asymmetry of $Z/\gamma^*$ + jets for SLT$_e$ is smaller than for SLT$_\mu$ because of the stricter requirements on the dielectron mass. 

Events due to QCD multijet production can enter the selection through hadronic misidentification or heavy-flavor decay. Missing transverse energy can arise from mismeasured jet energy, detector effects, or neutrinos in the decay chain. We estimate this background by releasing the $\slashed{E}_T$ requirement on the events and fitting templates of the $\slashed{E}_T$ distribution for the QCD multijet component, separately for OS and SS events. The template distribution for QCD multijet events is derived from a jet-enriched data sample in which candidate electrons fail two of the electron identification criteria. The remaining sample composition is modeled with MC simulations. 

Finally, the production of dibosons ($WW,WZ,ZZ$) and $t\bar{t}$ is modeled with a {\sc pythia} ({\sc v6.4}) MC calculation, while single top-quark production is simulated using {\sc madevent} \cite{madevent}. The $WW$ events contribute the most and have a strong charge asymmetry. Table~\ref{tab:summary2} summarizes the data and the estimated background. 
\begin{table}[th]
\caption{{\small Summary of data and backgrounds in the SLT$_\mu$-tagged and SLT$_e$-tagged $W$ + 1 jet samples. 
}}
\begin{center}
\begin{ruledtabular}
\begin{tabular}{lccc}
Source                & Events  &     Asymmetry            & OS--SS \\
\hline
SLT$_\mu$  &                        &       &  \\
\hline
$W$ + LF, $b\bar{b}$, $c\bar{c}$ & $1808\pm271$   & $0.048\pm0.008$  & $86\pm14$ \\
$Z/\gamma^*$ + jets  &    $132\pm30$      &   $0.63\pm0.02$   & $84\pm18$ \\
QCD multijet     & $308\pm17$    &  $-0.03\pm0.07$     &  $-8\pm17$ \\
Diboson, $t(\bar{t})$ &   $26\pm3$   &  $0.33\pm0.01$ & $9\pm1$ \\
$Wc$ (LO + PS)   & $214\pm19$    &   $0.75\pm0.03$  &  $161\pm13$ \\
\hline
Total expected     &  $2488\pm274$  & $...$ & $331\pm37$ \\
Data   &   2506     &   $...$   & $458$ \\
\hline
              &                        &       &  \\
  SLT$_e$       &                        &       &  \\
\hline
$W$ + LF, $b\bar{b}$, $c\bar{c}$  & $4076\pm305$   & $0.043\pm0.005$  & $174\pm19$ \\
$Z/\gamma^*$ + jets   &    $138\pm29$      &   $0.26\pm0.01$   & $36\pm7$ \\
QCD multijet     & $374\pm12$    &  $0.07\pm0.03$     &  $27\pm12$ \\
Diboson, $t(\bar{t})$  &   $35\pm3$   &  $0.58\pm0.01$ & $20\pm2$ \\
$Wc$ (LO + PS)   & $174\pm16$    &   $0.45\pm0.02$  &  $78\pm7$ \\
\hline
Total expected     &  $4797\pm307$  & $...$  & $336\pm28$ \\
Data   &   4582     &    $...$   & $406$ \\
\end{tabular}
\end{ruledtabular}
\end{center}
\label{tab:summary2}
\end{table}

\begin{figure}[th]
\begin{center}
\includegraphics[width=8cm]{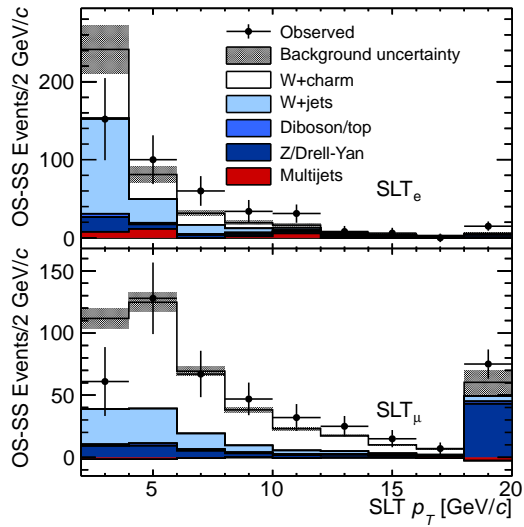}
\end{center}
\caption{{\small (color online) The soft muon and soft electron $p_T$ distributions. The $Wc$ contribution is normalized to the measured cross section.}}
\label{fig:kindist}
\end{figure}

We assume that the total OS--SS rates observed in the data, after subtracting the background contributions, are due to the $Wc$ signal; the SS-subtracted rates for the signal are then $287\pm 50({\rm stat})\pm 32({\rm syst})$ and $149\pm 68({\rm stat})\pm 26({\rm syst})$ events, for the SLT$_\mu$ and SLT$_e$ tagged samples, respectively. 
The total systematic uncertainty in the SS-subtracted rates is derived accounting for correlations between the uncertainties of the individual background sources. Figure \ref{fig:kindist} shows the distributions of the measured $p_T$ spectrum for SLT muons and electrons in tagged events, compared to the prediction. For each contribution, SS events are subtracted. The $Wc$ production cross section is calculated using Eq.~(\ref{eq:Xsec}), with $\sigma_{Wc}\equiv\sigma_{W^+\bar{c}}+\sigma_{W^-c}$, $B(W\rightarrow \ell\nu)=0.108\pm0.009$ \cite{pdg}, $p_{Tc}>20~{\rm GeV}/c$, and $|\eta_c|<1.5$; the values of the dilution $S$ for $Wc$ events are given in Table~\ref{tab:summary2}.We measure $\sigma_{Wc}\times B\left(W\rightarrow \ell\nu\right)=13.4\pm2.3({\rm stat})^{+2.5}_{-2.0}({\rm syst})^{+1.2}_{-1.0}({\rm lum})~{\rm pb}$ and $\sigma_{Wc}\times B\left(W\rightarrow \ell\nu\right)=15.0\pm6.8({\rm stat})^{+4.4}_{-2.9}({\rm syst})\pm1.2({\rm lum})~{\rm pb}$ from the SLT$_\mu$ and SLT$_e$ samples, respectively. 
\begin{table}[ht]
\caption{{\small Summary of systematic uncertainties, as a percentage of the measured $Wc$ cross section. Numbers shown in bold font indicate uncertainties treated as uncorrelated in the combination of the channels.
}}
\begin{center}
\begin{ruledtabular}
\begin{tabular}{lcc}
Source                  &  SLT$_\mu$ & SLT$_e$  \\
\hline
SLT uncertainties           &   $\pm${\bf 9.2}   & $\pm${\bf 16.6} \\
QCD multijet estimate       &   $\pm${\bf 6.3}   &  $\pm${\bf 9.9} \\
Initial and final state radiation &   $\pm6.0$   &  $\pm6.0$ \\
Background cross sections   &   $\pm5.7$   &  $\pm4.7$ \\
$c$ quark hadronization     &   $\pm4.6$   &  $\pm4.6$ \\
PDFs                        &   $\pm3.6$   &  $\pm3.6$ \\
$W$-lepton ID               &   $\pm2.2$     &  $\pm2.2$ \\
Jet energy calibration      &   $\pm2.0$   &  $\pm2.0$ \\
Factorization, renormalization scales         &   $\pm1.3$   &  $\pm1.3$ \\
\hline
Total			   &   $\pm$15.4  & $\pm$21.8 \\
\hline
Luminosity                  &   $\pm7.9$   & $\pm8.3$  \\
\end{tabular}
\end{ruledtabular}
\end{center}
\label{tab:summary3}
\end{table}

Systematic uncertainties are shown in Table~\ref{tab:summary3}. The uncertainty on the SLT tagging includes contributions from the measurements of the efficiency of tagging leptons in a jet environment and of mistagging \cite{smtprd2, setprd}. The uncertainty on the backgrounds includes contributions from the theoretical cross sections, from the estimation technique, and from statistics for the backgrounds evaluated with inputs from a data control region. For the $Z/\gamma^*$ background, the dominant uncertainty on the event yield estimate comes from the measured $Z$ cross section uncertainty. To measure the effects of initial- and final-state gluon radiation, we measure the $Wc$ acceptance in different samples with the radiation enhanced or reduced, as in Ref. \cite{topmass}. We compare charm jets modeled with the {\sc pythia} and {\sc herwig} \cite{herwig1,herwig2} MC calculations to evaluate the uncertainty due to different hadronization models. The PDF uncertainty is derived by remeasuring the acceptance using the CTEQ and MRST \cite{mrst} sets, following the same prescription as in Ref. \cite{topmass}. The MC modeling of the efficiency for identifying the leptons from the $W$ boson decay (``$W$ lepton ID") is measured using $Z$ boson data and MC samples. The charge misidentification rate is less than 1\% and therefore has a negligible effect. The uncertainty due to the jet energy calibration is measured by shifting the energies of the jets in the $Wc$ MC simulation by $\pm1\sigma$ of the jet energy calibration \cite{jes}. The uncertainty on the acceptance due to the factorization and renormalization scales is estimated by varying them in the {\sc alpgen} MC program between 1/2 and twice the transverse mass of the $W$ boson, as well as using the charm quark $p_T$. 

The results from the two SLT-tagged samples are combined by performing a profile likelihood ratio minimization \cite{cowan} in which the number of signal and background events in each sample is modeled by a Poisson distribution. Systematic uncertainties are included as nuisance parameters with Gaussian constraints whose widths are fixed to the respective uncertainties, and are assumed to be either fully correlated, if they are shared between the two channels, or uncorrelated if not. The cross section, $\sigma_{Wc}$, is left as a free parameter in the fit of the likelihood function. The combination yields $\sigma_{Wc}\left(p_{Tc}>20~{\rm GeV}/c,~|\eta_c|<1.5\right)\times B\left(W\rightarrow \ell\nu\right)= 13.6\pm2.2({\rm stat})_{-1.9}^{+2.3}({\rm syst})\pm1.1({\rm lum})~{\rm pb}=13.6_{-3.1}^{+3.4}$~pb. The significance for the $Wc$ signal is derived from the ratio of profile likelihoods $\lambda$, with $-2{\rm ln}\lambda$ in the hypothesis of no signal being interpreted as following a $\chi^2$ distribution, and is calculated to be 5.7$\sigma$. The measurement is in agreement with a NLO calculation over the same phase space of $11.4\pm1.3~{\rm pb}$ \cite{mcfm}, where the renormalization and factorization scales have been set to half the $W$ boson mass, and varied between 5 and 80~GeV in the uncertainty. The uncertainty also includes PDF variations using the {\sc CTEQ6M} \cite{cteq6m} and {\sc MSTW2008} \cite{mstw2008} sets. The result can be also compared to the LO prediction of $8.2\pm1.5$~pb \cite{mcfm}, giving a measurement to LO cross section ratio for this kinematic region of $1.6\pm0.5$. 
Since the majority of $Wc$ production proceeds through $c$ to $s$ quark coupling, we can relate the measured value of the cross section with the theoretical prediction and derive $|V_{cs}|$. Using $\sigma^{\rm theory}_{Wc}=9.8(\pm1.1)|V_{cs}|^2+2.1(\pm0.2)$~pb \cite{mcfm} we obtain $|V_{cs}|=1.08\pm0.16$, where the uncertainties in the cross section measurement and in the theoretical prediction have been added in quadrature. Restricting the range of $|V_{cs}|$ to the interval [0,1], a lower limit of $|V_{cs}|>0.71$ at the 95\% confidence level is extracted.

We thank the Fermilab staff and the technical staffs of the participating institutions for their vital contributions. This work was supported by the U.S. Department of Energy and National Science Foundation; the Italian Istituto Nazionale di Fisica Nucleare; the Ministry of Education, Culture, Sports, Science and Technology of Japan; the Natural Sciences and Engineering Research Council of Canada; the National Science Council of the Republic of China; the Swiss National Science Foundation; the A.P. Sloan Foundation; the Bundesministerium f\"ur Bildung und Forschung, Germany; the Korean World Class University Program, the National Research Foundation of Korea; the Science and Technology Facilities Council, the Royal Society and the Leverhulme Trust, U.K.; the Russian Foundation for Basic Research; the Ministerio de Ciencia e Innovaci\'{o}n, and Programa Consolider-Ingenio 2010, Spain; the Slovak R\&D Agency; the Academy of Finland; and the Australian Research Council (ARC).


\begin{thebibliography}{99.}
\bibitem{cdfsys} We use a ($z,\phi, \theta$) coordinate system with the $z$ axis in the direction of the proton beam, and $\phi$, $\theta$ the azimuthal and polar angles, respectively. Transverse energy and momentum are $E_T\equiv E\sin\theta$ and $p_T\equiv p\sin\theta$, respectively. The missing transverse energy is $\slashed{E}_T \equiv |-\sum_i E_T^i\hat{n}_i|$, where $E_T^i$ is the magnitude of the transverse energy contained in each calorimeter tower $i$ in the region $|\eta|<3.6$, and $\hat{n}_i$ is the direction unit vector of the tower in the plane transverse to the beam direction.
\bibitem{iso} The isolation ($I$) is defined as the calorimeter transverse energy in a cone of opening $\Delta R\equiv \sqrt{(\Delta\eta)^2+(\Delta\phi)^2}=0.4$ around the lepton (not including the lepton energy itself) divided by the lepton $E_T$ or $p_T$.
\bibitem{mtw} The transverse mass of the $W$ boson ($M_{T,W}$) is defined as $M_{T,W}=\sqrt{2p_{T,\ell}\slashed{E}_T(1-{\rm cos}\Delta\phi_{\ell,\nu})}$, where $\Delta\phi_{\ell,\nu}$ is the angle in the $r-\phi$ plane between the directions of $p_{T,\ell}$ and $\slashed{E}_T$.
\end{thebibliography}
\end{document}